# Domain-adaptive Fall Detection Using Deep Adversarial Training

Kai-Chun Liu, Michael Chan, Heng-Cheng Kuo, Chia-Yeh Hsieh, Hsiang-Yun Huang, Chia-Tai Chan and Yu Tsao

*Abstract*— Fall detection (FD) systems are important assistive technologies for healthcare that can detect emergency fall events and alert caregivers. However, it is not easy to obtain large-scale annotated fall events with various specifications of sensors or sensor positions during the implementation of accurate FD systems. Moreover, the knowledge obtained through machine learning has been restricted to tasks in the same domain. The mismatch between different domains might hinder the performance of FD systems. Cross-domain knowledge transfer is very beneficial for machine-learning based FD systems to train a reliable FD model with well-labeled data in new environments. In this study, we propose domain-adaptive fall detection (DAFD) using deep adversarial training (DAT) to tackle cross-domain problems, such as cross-position and cross-configuration. The proposed DAFD can transfer knowledge from the source domain to the target domain by minimizing the domain discrepancy to avoid mismatch problems. The experimental results show that the average F1-score improvement when using DAFD ranges from 1.5% to 7% in the cross-position scenario, and from 3.5% to 12% in the cross-configuration scenario, compared to using the conventional FD model without domain adaptation training. The results demonstrate that the proposed DAFD successfully helps to deal with cross-domain problems and to achieve better detection performance.

## I. INTRODUCTION

Falls are one of the main health risks to old people. The World Health Organization (WHO) has reported that about 28% of people over 65 years of age fall at least once each year [1]. The fallers might suffer from severe injuries or even death if no immediate medical aids are available. To provide timely intervention for fallers, fall detection (FD) alarm systems have become an important research topic in assistive technology and tele-healthcare. FD alarm systems with advanced wireless sensor networks and pattern recognition techniques have the capability to detect the occurrence of fall events in daily living and inform clinical professionals of emergency events. Furthermore, such alarm systems could alleviate the psychological stress of old people and caregivers [2].

Many studies have applied wearable inertial measurement units (IMU) to automatic FD systems [3-5] due to their various advantages, such as small size, low power, low cost, unobtrusiveness, and low invasivity. Accelerometers are the most popular measurement sensors in the field of wearable-based FD systems. They can measure the changes in body gestures and provide important movement information for FD systems.

FD approaches can be roughly divided into two categories, namely rule-based and machine-learning-based approaches. Rule-based approaches are designed based on human knowledge to develop hierarchical decision-making rules and thresholds for FD [6, 7]. However, the detection accuracy of rule-based FD models is general limited while dealing with experiments. To achieve better performance, several machine-learning-based FD systems [8] with advanced knowledge extractor and class discriminability have been proposed, involving support vector machine [8], k-nearest-neighbors [8], convolution neural networks [9], and recurrent neural networks [10]. Compared to rule-based approaches, these machine-learning-based FD systems require more efforts on parameter tuning or large data volume for model training.

Cross-domain problems in FD systems are still an open research topic. Among them, cross-position and cross-configuration problems are technical challenges in the field of wearable-based FD. Silva *et al.* [11] firstly explored the performance of FD systems on unseen sensor positions. They trained the FD model by fusing the data collected from multiple sensor positions. The results show that their model using multi-position can achieve satisfactory generalization for particular unseen positions. However, the impact of cross-position on detection performance has not been fully explored and only one public dataset is validated in their works. Delgado-Escaño *et al.* [12] developed a cross-dataset deep learning-based fall detector for tackling cross-configuration problems. They employed multi-task learning approaches and CNN-LSTM models to obtain effective features that can perform well in various datasets. However, their approaches required well-labeled data for the target domains.

FD systems require the ability to adapt to unseen environments, while it is challenging to collect data for all conditions in real-world applications. A typical solution [13] is to collect new data (including falls and ADL) from users during their daily living and ask them to label the data. Then, the FD model is fine-tuned based on the domain-specified data. However, manual labeling for the new data is prohibitively time-consuming and expensive. To tackle this challenge, unsupervised domain adaptation approaches have been developed to extract transferrable knowledge from the labeled dataset (source domain) for use in the unlabeled dataset (target domain) [14-17]. These approaches have achieved promising performance in numerous research fields of speech [18], computer vision [19], and natural language processing [20].

This work was supported in part by grants from the Ministry of Science and Technology, Taiwan, under Grant MOST 109-2221-E-001-022- and Grant 108-2628-E-001 -002 -MY3.

K.-C. Liu, M. Chan, H.-C. Kuo and Y. Tsao are with the Research Center for Information Technology Innovation (CITI) at Academia Sinica, Taipei 11529, Taiwan.
C.-Y. Hsieh, H.-Y. Huang, and C.-T. Chan are with the Department of Biomedical Engineering, National Yang Ming Chiao Tung University, Taipei 11221, Taiwan.
(Corresponding author: Y. Tsao, e-mail: yu.tsao@citi.sinica.edu.tw).

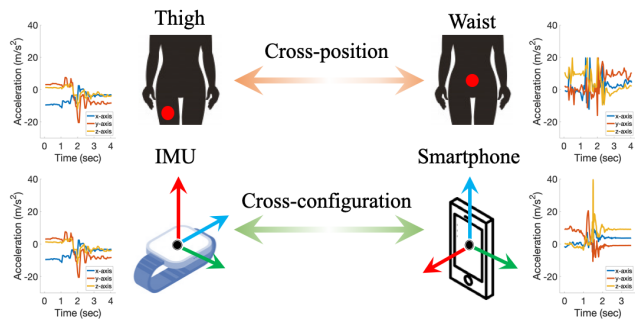

Fig. 1 Mismatch problems in real-world implementation of FD systems

One common domain adaptation approach is domain-adversarial training [21, 22], which aims to perform adaptive classification based on domain-invariant feature extraction. The main idea is to train a feature extractor that could enhance target label predictions and confuse domain classification.

As shown in Fig. 1, the present work aims to tackle two potential cross-domain circumstances when designing FD systems for real-world applications: (i) Cross-position: The previous study has shown that people have their preferable sensor positions according to individual preferences [23]. The detection algorithms designed for the single sensor position may limit the usability of the applications [24]. Therefore, the FD systems have to work well in multiple positions. Clearly, different sensor placements can cause diverse movement patterns. Moreover, it is difficult to collect the labeled data from all possible sensor positions. (ii) Cross-configuration: The distribution of the recorded patterns is heterogeneous under different hardware conditions (e.g., sampling rate, sensing range, resolution, and noise density), as various commercial sensors are available in the wearable device market [12]. For example, the model trained on the smartphone is unsuitable for a direct deployment on smartwatches because of the possible differences in sampling rate, sensing range, and resolutions. It is rather time-consuming to implement new fall experiments and data labeling for every unseen configuration.

In this work, a novel domain-adaptive fall detection (DAFD) system with a deep adversarial training (DAT) technique is proposed to tackle the domain discrepancy issue. The proposed DAFD system contains a CNN-based discriminative framework that can extract representative movement characteristics and critical fall features. By using the domain discriminator, the distribution of the extracted features from the source domain (i.e., waist and smartphone) aligns with that from the target domain (i.e., thigh and IMU). With DAT, the proposed DAFD could achieve better detection performance in the target domain.

The main contribution of this work is as follows:
- While most typical FD systems require extra labels for new domain adaptation, the proposed approach using unsupervised domain adaptation can adapt the detection model with unlabeled data of the target domain. It can greatly decrease the burden of data labeling and enhance the usability of the proposed FD model in real-world applications.
- We firstly propose a DAFD model that could transfer knowledge from the source domain (labeled data) to the target domain (unlabeled dataset) by minimizing the domain discrepancy for cross-position and cross-configuration problems. This work serves as the pioneering work that deals with cross-domain issues using DAT in wearable FD systems.

The remainder of this paper is organized as follows: In Section II, we introduce the chosen open datasets and their experimental setups. Section III presents the pre-processing processes of the accelerometer signals, involving resampling, impact-defined window, and min-max normalization. Section IV describes the architecture and design mechanisms of the proposed DAFD. The experimental results using the DAFD are presented in Section V. The comprehensive performance analysis of the proposed model for FD systems, its limitations, and future works are discussed in Section VI. Finally, we conclude this work in Section VII.

## II. OPEN DATASETS

Based on our survey results, more than ten datasets are publicly available for the evaluation of FD systems [25]. This study utilizes UP-Fall [26] and UMAFall [27] datasets to validate the proposed DAFD systems, while the objective is to tackle the technical challenges in cross-position and cross-configuration. Compared to other datasets, these two datasets place more than four sensor nodes on different locations of the body. Moreover, the employed sensor nodes and the configuration of these two datasets are completely different, showing that the datasets specially suit the main target of this study.

### A. UP-Fall

The UP-Fall dataset was published by Martínez-Villaseñor *et al.* [26]. The dataset consists of 17 healthy young subjects (9 males, 8 females, 18-24 years old, mean height: 1.66 m, mean weight: 66.8 kg) with five fall types and six types of activities of daily living (ADL). Due to data loss and overflow errors, 20% of the data were excluded. On average, 992 fall instances and 1197 ADL instances from accelerometers were used in this study. Details of the fall and ADL types are listed in TABLE I. Five tri-axial accelerometers with a sampling rate of 18.4 Hz were placed on the neck (N), waist (WA), right pocket (RP), wrist (WR), and ankle (A), as shown in Fig. 2.

### B. UMAFall

Another dataset, UMAFall, published by Santoyo-Ramón et al. [27] was used in this study. Four tri-axial accelerometers were placed on the chest (C), waist (WA), wrist (WR), and ankle (A), and a tri-axial accelerometer of the smartphone was placed in the left pocket (LP), as shown in Fig. 2. The accelerometers of IMUs and a smartphone were utilized to measure movement signals during the experiment. The sampling rates of the IMUs and the smartphone were 20 Hz and 200 Hz, respectively. The UMAFall dataset involved 19 subjects (11 males, 8 females, 19-68 years old, average height: 1.70 m, and average weight: 71.63 kg) for the experiments. However, seven of the subjects did not perform falls, and data loss and overflow errors were also found in this dataset. On average, only 740 fall instances and 1501 ADL instances collected from 12 subjects were kept, and about 20% of the UMAFall dataset was excluded. Finally, a total of 3 fall types and 11 ADL types were used in this dataset, as shown in TABLE II.

TABLE I
LIST OF FALL AND ADL IN UP-FALL

| NO. | Type of Fall and ADL | Instances |
|---|---|---|
| A1 | Walking | 249 |
| A2 | Standing | 249 |
| A3 | Sitting | 249 |
| A4 | Picking up an object | 249 |
| A5 | Jumping | 249 |
| A6 | Laying | 239 |
| F1 | Falling forward using hands | 249 |
| F2 | Falling forward using knees | 249 |
| F3 | Falling backwards | 249 |
| F4 | Falling sideward | 244 |
| F5 | Falling sitting in empty chair | 249 |

TABLE II
LIST OF FALL AND ADL IN UMAFALL

| NO. | Type of Fall and ADL | Instances |
|---|---|---|
| A1 | Applauding | 194 |
| A2 | Raising both arms | 200 |
| A3 | Emulating a phone call | 195 |
| A4 | Opening a door | 185 |
| A5 | Sitting on a chair and getting up | 86 |
| A6 | Walking | 246 |
| A7 | Bending | 228 |
| A8 | Hopping | 206 |
| A9 | Lying down on/standing up from a bed | 66 |
| A10 | Going upstairs and downstairs | 142 |
| A11 | Jogging | 129 |
| F1 | Forwards fall | 313 |
| F2 | Backwards fall | 321 |
| F3 | Lateral fall | 292 |

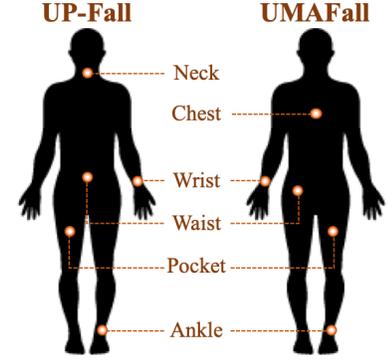

Fig. 2 An illustration of the sensor placements for UP-Fall and UMAFall datasets.

## III. PRE-PROCESSING

The resampling approach is first applied to the datasets as the sampling rates of UP-Fall and UMAFall are different. It is necessary to unify the dimension during the training and testing stages. In this work, the dataset with a higher sampling rate is downsampled to the sampling rate of the dataset with a lower sampling rate. Therefore, the UMAFall dataset with higher sampling rates (20 Hz and 200 Hz) were resampled to 18.4 Hz (sampling rate of UP-Fall dataset). A simple linear time warping resampling approach is employed in the UMAFall dataset. The UMAFall dataset is resampled by a resampling factor p/q, where p is a positive integer factor of interpolation (upsampling), and q is a positive integer factor of decimation (downsampling) [28]. Such an approach is implemented using Signal Processing Toolbox in the MATLAB 2016b environment. In this study, the resampling factor p/q is determined as 23/25 and 23/250 for the sampling rate of 20 Hz and 200 Hz, respectively.

Next, an impact-defined window is employed for processing the data. Such a windowing approach has shown superior detection performance in previous studies [29-32]. The main idea is to segment the data sequence based on the detection of the critical impact point during the entire fall event. The segmented data could cover several important states of the fall event, involving free-fall, vibration, and resting. The impact-defined window is determined as follows: First, the $Norm_{xyz}$ of the input three-dimensional accelerometer data $S = \{s_i | i = 1,2,\dots,n_S\}$ is calculated:

$$Norm_{xyz}(s_i) = \sqrt{a_{x_i}^2 + a_{y_i}^2 + a_{z_i}^2}, \quad (1)$$

where $n_S$ is defined as the total data samples of $S$ and $a_{x_i}, a_{y_i}$, and $a_{z_i}$ are the x-axis, y-axis, and z-axis samples of $s_i$. Then, the impact point $s_P$ is determined as the maximum value of $Norm_{xyz}$. Finally, the impact-defined window is determined as $W_p = \{s_{p-WS_b}, \dots, s_{p-1}, s_p, s_{p+1}, \dots, s_{p+WS_f}\}$, where $WS_f$ and $WS_b$ are the window sizes of the forward and backward sub-windows, respectively. In this study, $WS_f$ and $WS_b$ are determined as 1.5 s and 2 s because a previous study showed that FD systems with such window size achieve the highest detection performance [30]. Therefore, the input is 66 (37+1+28) data samples with a sampling rate of 18.4 Hz.

After the windowing process, a min-max normalization is used to reduce scaling effects on the DAFD model during the training phase. Assume that a series of data $S = \{s_j | j = 1,2,\dots,n_S\}$, where $s_j$ can be normalized to [0,1] with min-max normalization:

$$s_j^{nom} = \frac{s_i - s_{min}}{s_{max} - s_{min}}, \quad (2)$$

where $s_{max}$ and $s_{min}$ are the maximum and minimum of $S$. In this work, the min-max normalization is applied to individual dimension of tri-axial accelerometers.

## IV. DOMAIN-ADAPTIVE FALL DETECTION

### A. Model Architecture

Given that labeled data segments $R_s = \{r_i^s, l_i | i = 1, 2, \dots, n_s\}$ collected from the source accelerometer are sufficient to perform task learning $T_s$ for FD in the source domain, where $r_i^s$ is the $i$th data segment, and $l_i \in \{ADL, Fall\}$ indicates the class of the $i$th data segment. Unlabeled data segments collected from the target accelerometer are denoted as $R_t = \{(r_j^t | j = 1, 2, \dots, n_t\}$, where $r_j^t$ is the $j$th data segment and $n_t$ is the total data sample of $R_t$. The main goal of domain adversarial training is to perform task learning $T_t$ for FD on unlabeled target data by extracting invariant knowledge from $T_s$ and $R_s$.

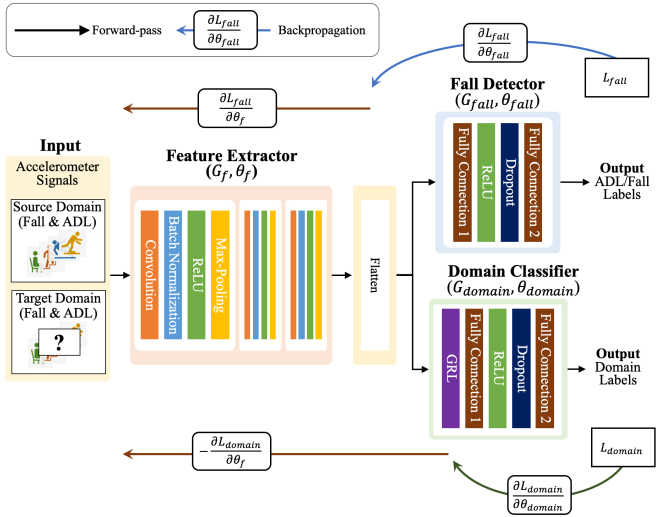

Fig. 3 The proposed adversarial training framework for the domain-adaptive fall detection (DAFD).

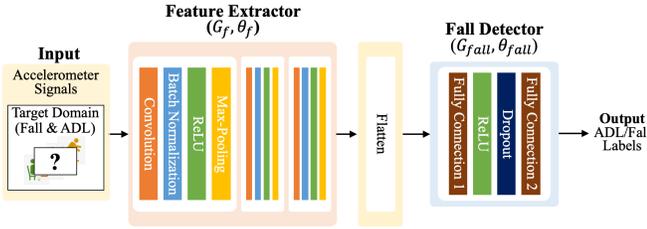

Fig. 4 The testing procedure for the proposed DAFD

In this study, domain-adversarial neural network (DANN) [22] is applied to the proposed DAFD that incorporates a domain classifier and a fall detector. The train and testing phases of the proposed DAFD using DANN are shown in Fig. 3 and Fig. 4, respectively. DANN is shown to generalize well from one domain to another while preserving low risk in the source domain in sentiment analysis and image classification tasks [22]. During the training phase, the fall detector learns the discriminative ability to identify falls and ADLs from $T_s$ and $R_s$. The domain classifier contains a gradient reversal layer (GRL), which keeps the input unchanged in the forward pass and reverses the gradient during backpropagation. Such operation allows extracting domain-invariant and discriminative features at the same time. In the testing phases, the trained fall detector is utilized to classify the ADL and fall data of the target domain.

The forward and backward propagation of GRL are defined as equation (4) and (5), respectively, where $R(r)$ is the function of GRL, $r$ is the input, and $I$ is the identity matrix.

$$R(r) = r, \quad (4)$$

$$\frac{dR}{dr_i} = -I, \quad (5)$$

The proposed DAFD consists of three components: a feature extractor $G_f$, a fall detector $G_{fall}$, and a domain classifier $G_{domain}$. The dimensions of the inputs and outputs of the models are as follows:

- Input of a 3D matrix has a dimension of 66 x 3.
- Features extracted by $G_f$ has a dimension of 10 x 4.
- Output of $G_{fall}$ has a dimension of 2.
- Output of $G_{domain}$ has a dimension of 2.

Each of the three components has its respective purpose:

- $G_f$ (input; $\theta_f$) learns a function that maps the input to the feature space.
- $G_{fall}$ (features; $\theta_{fall}$) learns a function that maps the features to the fall/ADL output space.
- $G_{domain}$ (features; $\theta_{domain}$) learns a function that maps the features to the domain output space.

$G_f$ takes a batch of tri-axial accelerometer data with 66 samples per axis. $G_f$ consists of three mixed layers. Each mixed layer comprises a convolutional layer (channel number = 4, kernel = 3, stride = 1), a batch normalization layer, a ReLU layer, and a max-pooling layer (kernel = 2, stride = 2). $G_f$ reduces the sample size from 66 to 10 and extracts high-level features. The four channels of extracted features are then flattened into a 1D vector with a dimension of 40. Next, the extracted features are fed into $G_{fall}$, which includes the first fully connected layer with a dimension of 40 x 50, a ReLU layer, a dropout layer, and the second fully connected layer with a dimension of 50 x 2. While the extracted features are also fed in parallel into $G_{domain}$, $G_{domain}$ is nearly identical to $G_{fall}$ except its first layer, which is GRL. Both outputs of $G_{fall}$ and $G_{domain}$ are converted to probability measures through the Softmax function and finally used to compute their cross-entropy loss. Both activity labels and domain labels are binary, representing fall/ADL and source/target, respectively.

### B. Parameter Training and Optimization

During the training phase, $L_{total}$ is a summation of $L_{fall}$ and $L_{domain}$, which in turn corresponds to the parallelism of $G_{fall}$ and $G_{domain}$. The loss function is formulated as follows:

$$loss_{total} = loss_{fall} + loss_{domain}, \quad (6)$$

$$loss_{fall} = \frac{1}{n_s}\sum_{i=1}^{n_s} L_{fall}\left(G_{fall}(G_f(r_i^s; \theta_f); \theta_{fall})\right), \quad (7)$$

$$loss_{domain} = \\ -\lambda\left[\frac{1}{n_s}\sum_{i=1}^{n_s} L_{domain}\left(G_{domain}\left(R(G_f(r_i^s; \theta_f); \theta_{domain})\right)\right) + \frac{1}{n_t}\sum_{j=1}^{n_t} L_{domain}\left(G_{domain}\left(R(G_f(r_j^t; \theta_f); \theta_{domain})\right)\right)\right], \quad (8)$$

where $L_{fall}$ and $L_{domain}$ are the negative log-probabilities of the correct label. Equation (6)-(8) enable the system to learn domain-invariant and FD features by maximizing $\theta_f$ and $\theta_{fall}$ and minimize $\theta_{domain}$.

Backpropagation is performed using batch gradient descent (batch size = 4 for each domain). The Adam optimizer [33] with an L2 regularization weight decay of 0.01 is used to update the parameter with a range of learning rates (the details are introduced in Subsection C). The update rule for parameters $\theta_{fall}, \theta_{domain}$, and $\theta_f$ is described as follows:

TABLE III
SEARCH RANGE FOR HYPERPARAMETER OPTIMIZATION

| Hyperparameter | Search range |
| --- | --- |
| Dropout | [0.1, 0.2, 0.5] |
| Learning rate | [0.001, 0.0005, 0.0001] |
| Domain regularization parameter ($\lambda$) | [0.31, 1, 1.3] |

TABLE IV
LIST OF TARGET-SOURCE PAIRS IN CROSS-POSITION AND CROSS-CONFIG. SCENARIOS

| Cross-position within Dataset (Source → Target) | | | | Cross-config. (Source → Target) | |
| --- | --- | --- | --- | --- | --- |
| UP-Fall | | UMAFall | | UP-Fall → UMAFall | UMAFall → UP-Fall |
| N→WA | RP→WR | C→WA | WR→A | WA→WA | WA → WA |
| N→LP | RP→N | C→WR | WR→WA | RP→LP | LP →RP |
| N→WR | WR→N | C→A | WR→C | WR →WR | WR →WR |
| N→A | WR→RP | WA→WR | A→C | A →A | A → A |
| WA→P | WR→A | WA→A | A→WA | | |
| WA→WR | WR→WA | WA→C | A→WR | | |
| WA→A | A→WA | | | | |
| WA→N | A→N | | | | |
| RP→A | A→RP | | | | |
| RP→WA | A→WR | | | | |

Note: **N**: neck, **WA**: waist, **RP**: right pocket, **LP**: left pocket, **WR**: wrist, **A**: ankle, **C**: chest

$$\Delta\theta_{fall} = -lr * \frac{dloss_{fall}}{d\theta_{fall}}, \theta_{fall} \leftarrow \theta_{fall} + \Delta\theta_{fall}, \quad (9)$$

$$\Delta\theta_{domain} = -lr * \frac{dloss_{domain}}{d\theta_{domain}}, \theta_{domain} \leftarrow \theta_{domain} + \Delta\theta_{domain}, \quad (10)$$

$$\Delta\theta_f = -lr(\frac{dloss_{fall}}{d\theta_{fall}}\frac{d\theta_{fall}}{d\theta_f} - \lambda\frac{dloss_{domain}}{d\theta_{domain}}\frac{d\theta_{domain}}{d\theta_f}), \theta_f \leftarrow \theta_f + \Delta\theta_f, (11)$$

where $lr$ is the learning rate and $\lambda$ is the domain regularization parameter.

Note that with this update rule, $G_{fall}$ and $G_{domain}$ would approach the local minimum with respect to their losses. While $loss_{fall}$ is backpropagated to update $\theta_f$, $loss_{domain}$ is backpropagated with an additional negative sign (an operation dictated by the GRL). $\theta_f$ is updated to reach a local maximum of $loss_{domain}$. Such adversarial training strengthens the discriminability of $G_{fall}$ and $G_{domain}$, and strengthens the fall/ADL discriminability, yet weakens the domain discriminability of the features extracted by $G_f$.

### C. Training and Validation Procedure

We tested performance with three training modes: "Source-only," "DAFD," and "Target-only." For "Source-only," only labeled source data is used to train the parameters of $G_f$ and $G_{fall}$. $loss_{fall}$ alone is used during backpropagation, so the model is unaffected by $loss_{domain}$. For "DAFD," both labeled source data and unlabeled target data are used to train the parameters of $G_f$, $G_{fall}$, and $G_{domain}$. Both $loss_{fall}$ and $loss_{domain}$ are used during backpropagation. For "Target-only," only labeled target data is used to train the parameters of $G_f$ and $G_{fall}$. The same as the "Source-only" setup, the "Target-only" uses $loss_{fall}$ alone, and thus the model is not affected by $loss_{domain}$. In this work, "Source-only" and "Target-only" are considered as the lower-bound and upper-bound detection performances to compare with the proposed DAFD. Because both datasets are imbalanced, we implemented weighted sampling to ensure that the model has seen roughly the same amount of fall and ADL samples. Specifically, the underrepresented fall data are oversampled to reach such a goal.

In this study, a leave-subjects-out cross-validation (LSOCO) approach is applied to validate the proposed DAFD. All subjects are divided into 5 groups and repeated five times to evaluate system performance on new subjects. Such an approach can achieve a more unbiased environments on the systems [34]. Furthermore, the proposed DAFD employs hyperparameter optimization and early stopping to prevent overfitting.

### D. Hyperparameter Optimization and Early Stopping

The three hyperparameters selected for fine-tuning are dropout, learning rate, and domain regularization parameter ($\lambda$). Three values are picked for each hyperparameter, accordingly for mining 27 permutations in total, as shown in TABLE III. The tuple with the lowest $loss_{total}$ is selected as the optimal hyperparameter. In addition to hyperparameter fine-tuning, early stopping is applied to prevent overfitting and to increase training efficiency. The criterion for early stopping is the epoch with the lowest $loss_{total}$, as shown in Eq. (6).

The proposed model was implemented using PyTorch 1.3.1, running on a workstation with 64-bit Ubuntu 18.04.4, Intel(R) Xeon(R) CPU E5-2683 v3 @ 2.00 GHz, and trained and tested using the Nvidia Titan Xp with 128 GB dedicated memory. The implementation of pre-processing was run in an MATLAB 2018 environment.

### E. Cross-domain Scenario

In this study, we set up two cross-domain scenarios for the experiments. The first is cross-position, which aims to explore the capability of the proposed FD model to align data distributions collected at different body positions in the same configuration. In total, there are 20 and 12 source-target pairs for the UP-Fall and UMAFall datasets, respectively. Note that the sensor placed in the left pocket in UMAFall data is excluded in this scenario because its configuration is different from other positions. The second is cross-configuration, which explores the detection ability of the proposed DAFD to align data distributions of UMAFall and UP-Fall datasets at each sensor location. The sensors placed on the neck and chest only occur in one of the datasets, while other sensor placements overlap. Therefore, the data collected from the neck and chest sensors excluded from the source-target pairs in cross-configuration scenarios. As a result, a total of 8 source-target pairs are formed in this scenario. All considered source-target pairs in cross-position and cross-configuration scenarios are listed in TABLE IV.

### F. Evaluation Methodology

Several evaluation metrics were used to validate the reliability of the proposed DAFD: sensitivity (SEN), specificity (SPE), precision (PRE), and F1-score (F1). These metrics are popular performance measures in FD systems. They are defined as follows:

$$SEN = \frac{TP}{TP+FN}, \quad (12)$$

TABLE V
OVERALL PERFORMANCE COMPARISON IN CROSS-POSITION AND CROSS-CONFIGURATION SCENARIOS (%)

|  |  | Cross-position | | Cross-config. (Source → Target) | |
|---|---|---|---|---|---|
|  |  | UP-Fall | UMA-Fall | UP-Fall → UMAFall | UMAFall → UP-Fall |
| SEN | Source-only | 86.62 | 62.72 | 85.63 | 53.06 |
|  | DAFD_adl | 88.48* | 62.78 | 78.68* | 57.17* |
|  | DAFD | 89.28† | 72.08† | 92.09† | 67.12† |
|  | Target-only | 94.68 | 92.07 | 95.05 | 97.23 |
| SPE | Source-only | 97.32 | 95.34 | 94.21 | 91.57 |
|  | DAFD_adl | 98.64 | 95.68 | 94.79 | 92.16 |
|  | DAFD | 97.25 | 94.65 | 93.78 | 94.56† |
|  | Target-only | 98.81 | 95.74 | 98.72 | 99.36 |
| PRE | Source-only | 96.58 | 89.31 | 89.37 | 86.49 |
|  | DAFD_adl | 98.49* | 89.53 | 87.43* | 87.20 |
|  | DAFD | 96.55 | 91.61† | 90.19 | 91.78† |
|  | Target-only | 98.56 | 94.06 | 97.96 | 99.22 |
| F1 score | Source-only | 91.24 | 72.31 | 87.23 | 65.63 |
|  | DAFD_adl | 92.98 | 72.45 | 82.83* | 69.31* |
|  | DAFD | 92.74 | 80.07† | 91.13† | 77.52† |
|  | Target-only | 96.57 | 92.94 | 96.43 | 98.20 |

* indicates significant differences ($p < 0.05$) between "Source-only" and DAFD_adl
† indicates significant differences ($p < 0.05$) between "Source only" and DAFD

$$SPE = \frac{TN}{TN+FP}, \quad (13)$$

$$PRE = \frac{TP}{TP+FP}, \quad (14)$$

$$F1 = \frac{2*SEN*PRE}{SEN+PRE}, \quad (15)$$

where TP, TN, FP and FN are the true positive, true negative, false positive, and false negative of the labels, respectively.

The significance of performance between different approaches was statistically measure using independent t-test at 95% ($p < 0.05$).

## V. EXPERIMENTAL RESULTS

Overall performance comparison in cross-position and cross-configuration scenarios are shown in TABLE V. The experimental results demonstrate that the detection performance of the proposed DAFD system is better than that of the "Source-only". The improvement in all SEN and F1 ranges from 1.5% to 14%, while the SPE and PRE of several pairs remain unchanged or slightly decrease (less than 1%). Such results show that the performance of the proposed DAFD and "Source-only" is significantly different ($p < 0.05$) in SEN and F1. However, the results still show that the performance of the "Target-only" is better than the proposed DAFD system. The gaps ranged from 1.22% to 30.11% for SEN and from 3.79% to 20.68% for F1.

An additional experiment is implemented to investigate the feasibility of the DAFD in unseen environments if the target data only involve ADL signals. The proposed DAFD model with ADL of target data only (DAFD_adl) is shown in TABLE V. The results show that DAFD_adl performs well in most scenarios but has limited capability in some particular scenarios, such as "UP-Fall→UMAFALL".

Fig. 5 and 6 show the detailed results of different source-target pairs in cross-position scenarios. The best and the worst F1-score of UP-Fall are WA→P (97.24%) and WR→A (84.63%), respectively, and those of UMAFall are C→WA (95.93%) and A→WR (60.48%), respectively. The results reveal that most positions close to the human body center, such as chest, neck, waist and pocket, can effectively adapt to each other to improve detection performance. However, the proposed DAFD fails in several source-target pairs related to the ankle and wrist, including WA→A, WA→WR, WR→P, WR→A for UP-FALL, and C→A, WA→A, and A→C for UMAFall.

Fig. 7 demonstrates the detection performance in the cross-configuration scenarios. Generally, the F1 of all source-target pairs is improved with the proposed DAFD, where the improvements range from 0.75% to 24.95%. Similar to the cross-position scenario, the proposed DAFD achieves better SEN performance as the improvement of SPE and PRE are relatively limited.

## VI. DISCUSSION

In this work, DAFD using DAT is proposed to deal with realistic cross-domain problems, e.g., cross-position and cross-configuration. The overall improvements in F1 and SEN range from 1.5% to 14%. The experimental results show that the proposed approach effectively tackles the problems and achieves better detection performance, especially for sensitivity.

As shown in Fig. 8, we use t-SNE [35] to visualize the distribution of the features extracted from the feature extractor $G_f$. Fig. 8 (a) shows that using the "Source-only" training model fails to identify fall/ADL data samples of the target domain because the distribution of the target domain is quite different from that of the source domain. In contrast, the FD model using DAT alleviates the discrepancy between the source and target domains, as shown in Fig. 8 (b). The proposed DAFD provides a clearer decision boundary between fall and ADL data points to the target distribution. Such results show that the proposed DAFD successfully utilizes the knowledge from the source domain to handle the mismatch issue.

Similar DAT approaches have been successfully implemented in other fields of applications, such as image processing [19], speech processing [18] and wearable-based ADL monitoring [36]. To the best of our knowledge, the feasibility of DAT in FD systems has not been demonstrated in previous work. The main contribution of this study is that we aim to tackle cross-domain problems using DAT in wearable FD systems, while most studies focus on improving detection performance in relatively ideal environments.

Several challenges remain and limit the detection ability of the proposed DAFD. First, the patterns collected from the sensors placed on the ankle and wrist are completely different from those on the neck, chest, waist, and pocket. It not only constrains the effectiveness of DAFD but also leads to worse performance in several source-target pairs. Second, it is challenging to train deep neural networks (DNN) with imbalanced data between fall and ADL and limited data volume from the public datasets. These problems often cause overfitting

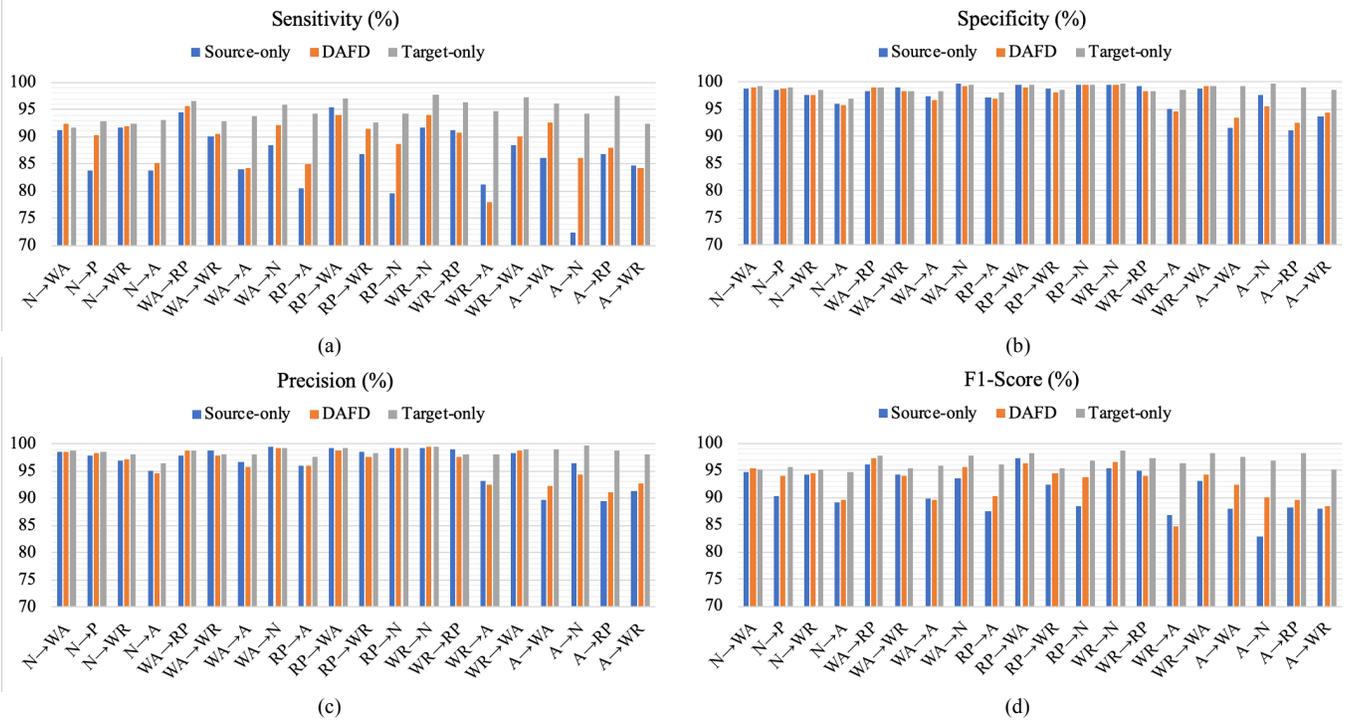

Fig. 5 Detection performance vs cross-position within UP-Fall dataset (a) sensitivity (b) specificity (c) precision (d) F1-score

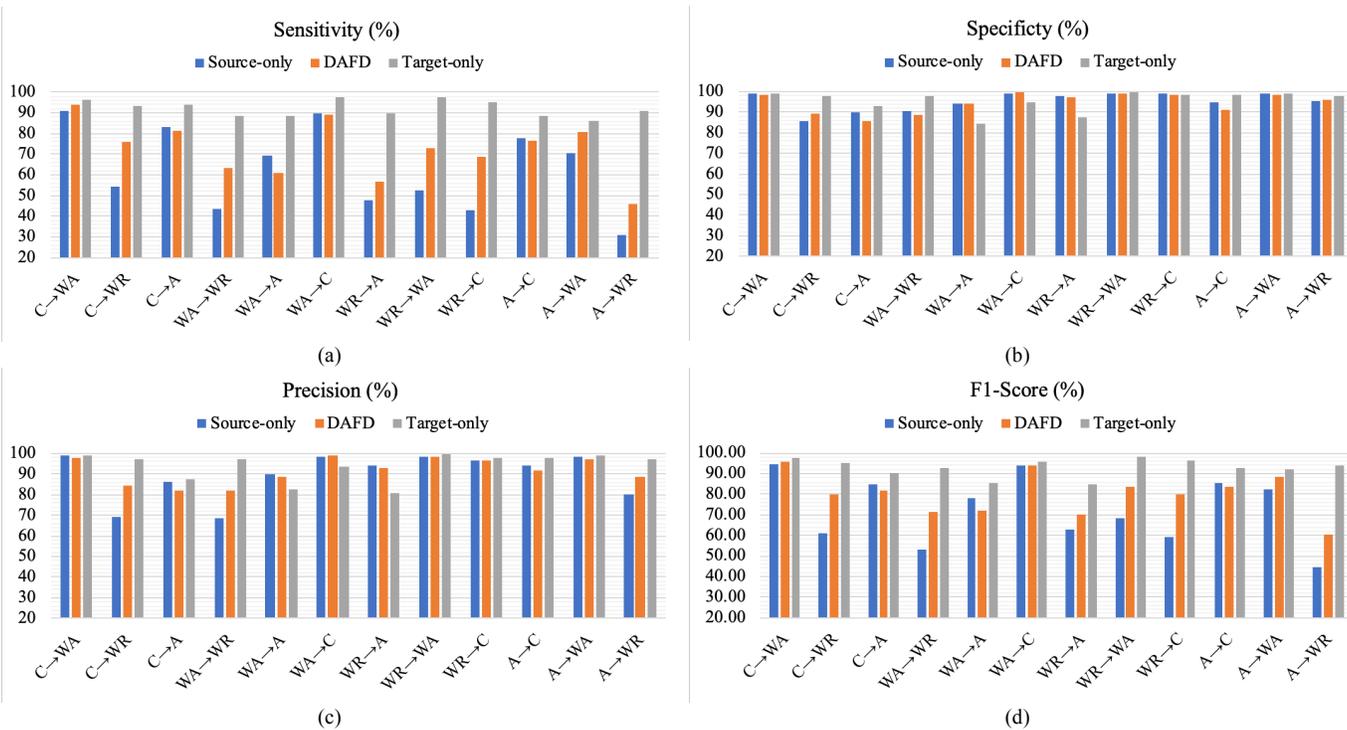

Fig. 6 Detection performance vs cross-position within UMAFall dataset (a) sensitivity (b) specificity (c) precision (d) F1-score

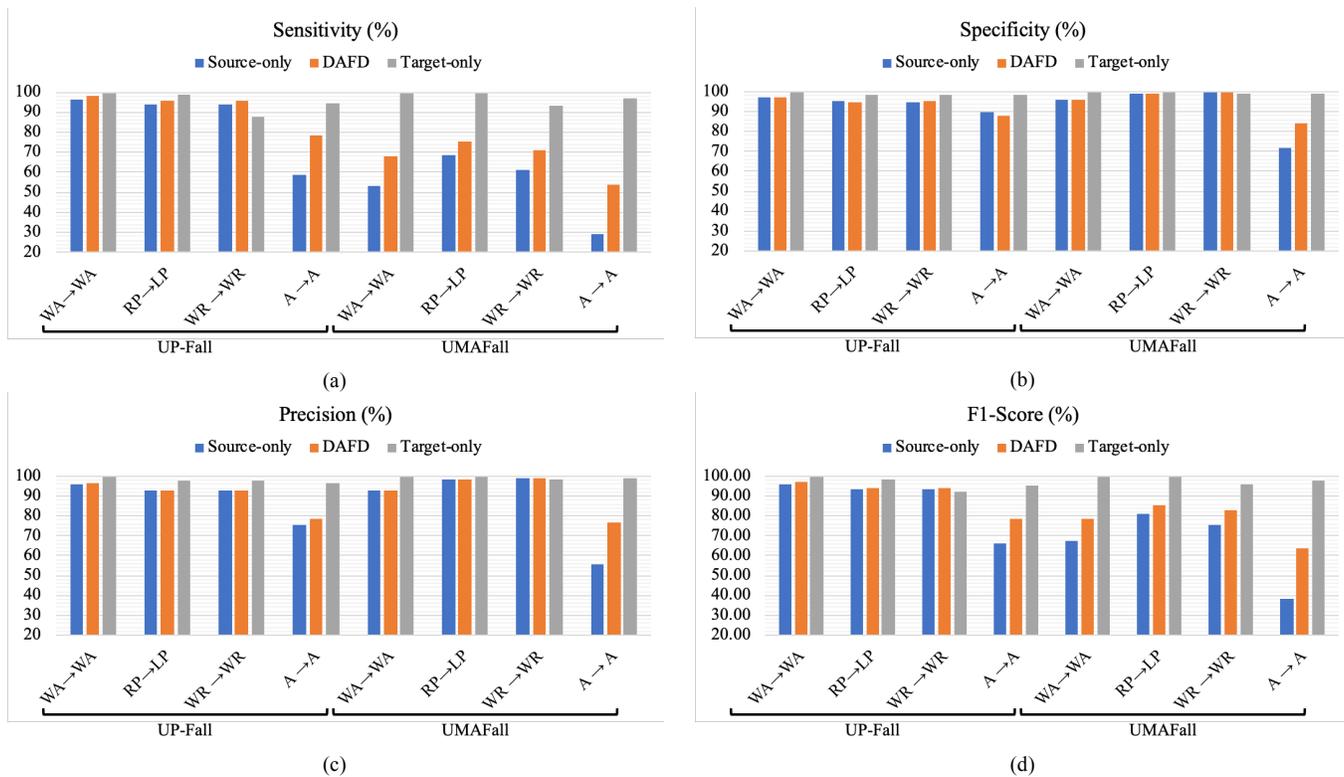

Fig. 7 Detection performance vs cross-config. (a) sensitivity (b) specificity (c) precision (d) F1-score

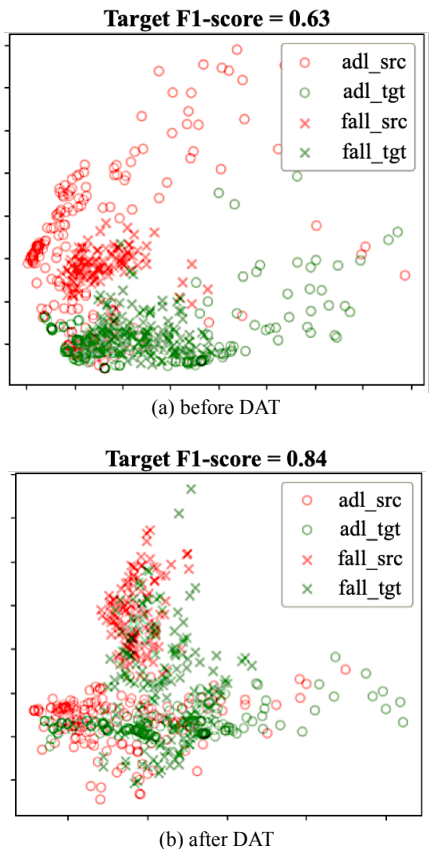

(a) before DAT

(b) after DAT

Fig. 8 Distribution of extracted features from NN using t-SNE visualization. Red and green points, respectively, represent source and target domains, while cross and circle correspond to fall and ADL. (a) NN trained with "Source-only" (without DAT) (b) using the proposed DAFD

problems in the proposed FD model. Therefore, DAFD is only implemented with few hidden layers and a small number of parameters in each layer. To further improve the system performance, advanced techniques and models will be employed for DAT, involving data augmentation [37], multi-task learning [38], and few-shot learning [39].

This study has shown that the proposed DAFD can adapt the FD model to the new environment with the unlabeled data of ADL and fall signals. Such an approach can greatly decrease the burden of manual labeling in the typical domain adaptation approach [13]. However, before FD systems go online, asking subjects performing fall events to prepare the target data for the domain adaptation may be not acceptable for users. A more suitable approach is to ask subjects performing ADL (without FD data) only as the target domain data. Therefore, we explore the detection performance using DAFD_adl that involved ADL signals of target data only (without FD data), and report the results in TABLE V. Unfraternally, the FD systems using DAFD_adl cannot perform well in such scenario. Please note that since only ADL data is available, this scenario is more related to FD system calibrations rather than domain adaptations. In future work, more advanced model designs and training procedures will be investigated for calibration requirements.

One limitation of this study is that we only employed the simulated datasets instead of real-world datasets (e.g., FARSEEING [40]). This is because the volume of the real-world dataset is too small to train DNN models. Obviously, it requires other pre-processing techniques before using DAT, such as data augmentation. Additionally, several mismatch factors are not considered in this study, involving cross-gender, cross-age (e.g., young adults vs. elderly) and cross-sensors (e.g., accelerometer vs. gyroscope). In our future study, we plan to explore and investigate the feasibility of DAT in these cross-domain scenarios.

VII. CONCLUSION

In this study, we propose DAFD that is based on the DAT criterion to deal with two potential cross-domain scenarios, cross-position and cross-configuration, when designing FD systems for real-world applications. The proposed model could transfer knowledge from the source domain to the target domain by minimizing the domain discrepancy while enabling extracted features to maintin discrimination fro the main learning task. In this study, the overall detection performance using DAFD outperforms that using typical FD systems by 1.5% to 14% for SEN and F1, repsectively. The experimental results demonstrated the feasibility and effectiveness of the proposed DAFD for tackling cross-domain problems.